\documentclass[journal]{IEEEtran}
\usepackage{amsmath,bm}
\usepackage[dvips]{graphicx}

\begin{document}

\title{Periodic and Quasi-Periodic Compensation Strategies
of Extreme Outages caused by Polarization Mode
Dispersion and Amplifier Noise}

\author{Vladimir Chernyak, Michael Chertkov, Igor Kolokolov, and Vladimir Lebedev
\thanks{ This work was supported by LDRD ER on ``Statistical Physics of Fiber Optics Communications"
at Los Alamos National Laboratory.}
\thanks{V. Chernyak is with Corning Inc., SP-DV-02-8, Corning, NY 14831, USA}
\thanks{M. Chertkov, V. Lebedev and I. Kolokolov are with
Theoretical Division, LANL, Los Alamos, NM 87545, USA}
\thanks{V. Lebedev and I. Kolokolov are also with
Landau Institute for Theoretical Physics, Moscow, Kosygina 2, 117334, Russia}
\thanks{I. Kolokolov is also with Budker
Institute of Nuclear Physics, Novosibirsk 630090, Russia.}}

\maketitle


\begin{abstract}

  Effect of birefringent disorder on the Bit Error Rate (BER) in
  an optical fiber telecommunication system subject to amplifier noise may lead to extreme
  outages, related to anomalously large values of BER. We
  analyze the Probability Distribution Function (PDF) of BER for
  various strategies of Polarization Mode Dispersion (PMD)
  compensation. A compensation method is proposed that is capable of more efficient
  extreme outages suppression, which leads to substantial improvement
  of the fiber system performance.

\end{abstract}




\IEEEpeerreviewmaketitle

Polarization Mode Dispersion (PMD) is an essential impairment for modern optical fiber systems
\cite{97PN,98Hei,00GK}. Therefore, dynamical PMD compensation became an important subject in
modern communication technology \cite{94OYSE,98HFW,99MK,00BBBBW}. Optical noise generated in
optical amplifiers represents another impairment that may not be reduced/compensated and,
therefore, should be also considered in any evaluation of a fiber system performance \cite{94Des}.
BER calculated for a given realization of birefringent disorder by means of averaging over the
amplifier noise statistics constitutes an appropriate object to characterize joint effect of the
two impairments. In two preceding papers \cite{03CCKLb,03CCKLc} we have demonstrated that the
probability of extreme outages (values of BER much higher than typical) is substantially larger
than one could expect from naive Gaussian estimates singling out effects of either of the two
impairments. The natural object of interest is the PDF of BER and, specifically, the PDF tail
corresponding to anomalously large BER. In \cite{03CCKLb} we have developed a consistent
theoretical approach to calculating this tail. The case when no compensation is applied and also
the effect of the simplest ``setting the clock" compensation on the PDF tail suppression have been
discussed in \cite{03CCKLb}. Then our investigation was extended to study effects of the standard
first- and higher- order compensations on extreme outages \cite{03CCKLc}. In the present letter we
propose a compensation scheme that appears to be more efficient in reducing the extreme outages
compared to the traditional high-order compensation scheme with the same number of the
compensating degrees of freedom.

We consider the return-to-zero (RZ) modulation format, when
optical pulses are well separated in time $t$, and thus can be
analyzed as individual objects. We represent the pulse intensity
measured at the system output as
 \begin{equation}
 I=\int \mathrm dt\, G(t)
 \left|{\cal K}\bm\varPsi(t)\right|^2 \,,
 \label{nnn} \end{equation}
where $G(t)$ is a convolution of the electrical (current) filter function with the sampling window
function. The two-component complex field $\bm\varPsi(t)$ describes the output optical signal (the
components correspond to two polarization states of the signal). The linear operator ${\cal K}$ in
Eq. (\ref{nnn}) represents optical filtering, it may also account for a compensating device. The
compensating part of the linear operator, ${\cal K}_c$, is applied first, i.e. before filtering
described by ${\cal K}_f$, resulting in ${\cal K}={\cal K}_f{\cal K}_c$. Ideally, $I$ takes two
distinct values depending on whether the information slot is vacant or filled. However, the
impairments enforce deviations of $I$ from those fixed values. If the output signal intensity
exceeds the decision level $I_d$, then ``1" is associated with the slot, otherwise the slot is
labeled by ``0". Sometimes the information is lost, i.e. the initial ``1" is detected as ``0" at
the output or vise versa. The BER is the probability of such events that naturally depends on a
specific realization of birefringent disorder in the fiber. BER must be extremely small to
guarantee successful system performance. It has been demonstrated in \cite{03CCKLb} that
anomalously high BER originates solely from the ``$1\to0$" events. We denote the probability of
such events by $B$ and study its sensitivity with respect to disorder.

In this letter we restrict ourselves to the linear propagation regime.
when the output signal $\bm\varPsi(t)$ can be represented as a sum of two
contributions: $\bm\varphi$, related to the noiseless evolution of the initial pulse, and the
noise-induced part $\bm\phi$. We consider the cases of distributed or, alternatively, lumped
amplification with the fiber length $Z$ essentially exceeding the inter-amplifier separation (span
length) within the same framework. $\bm\phi$ becomes a zero-mean Gaussian variable,
completely characterized by its two-point correlation function
 \begin{equation}
 \langle\phi_\alpha(t_1)\phi^\ast_\beta(t_2)\rangle
 =D_\xi Z\delta_{\alpha\beta} \delta(t_1-t_2),
 \label{phiphi} \end{equation}
insensitive to a particular realizations of birefriengent disorder and chromatic dispersion
in the fiber. The product $D_\xi Z$ represents the amplified spontaneous emission (ASE) spectral
density accumulated along the fiber.
The coefficient $D_\xi$ is introduced into Eq. (\ref{phiphi}) to reveal the
linear growth of the ASE factor with $Z$ \cite{94Des}. The noise-independent part of the
signal is governed by
 \begin{eqnarray}
 \partial_z{\bm\varphi}
 -\hat{m}(z)\partial_t{\bm\varphi}
 -id(z)\partial_t^2{\bm\varphi}=0,
 \label{varphi} \end{eqnarray}
$z$ and $d$ being the coordinate along the fiber and chromatic dispersion.
The birefringence matrix can be represented as $\hat m=h_j\hat{\sigma}_j$, where $h_j$ is
a real three-component field and $\hat{\sigma}_j$ are the Pauli matrices. Averaging over many
states of the birefriengent disorder, any fiber is going through over time, or (alternatively)
over the states of birefringence of different fibers, one finds that $h_j(z)$ is a zero-mean
Gaussian field described by
 \begin{equation}
 \langle h_i(z_1)h_j(z_2)\rangle
 =D_m\delta_{ij}\delta(z_1-z_2).
 \label{hh} \end{equation}
If birefringent disorder is weak the integral $\bm H=\int_0^Z\mathrm dz\,\bm h(z)$
coincides with the PMD vector. Thus, $D_m=k^2/12$, where $k$ is the so-called PMD coefficient.

In an operable communication system, typical damage caused by disorder and noise must be small,
i.e. typically both impairments can cause only a small distortion to a pulse, thus, the optical
signal-to-noise ratio (OSNR) and the ratio of the squared pulse width to the mean squared value of
the PMD vector are both large. OSRN can be estimated as $I_0/(D_\xi Z)$ where $I_0=\int\mathrm
dt\,|\varPsi_0(t)|^2$ is the initial pulse intensity, the integration being performed over a
single slot populated by an ideal pulse, encoding ``1''. Typically, $B$ fluctuates around $B_0$,
the zero-disorder ($h_j=0$) value of $B$. A convenient auxiliary dimensionless object,
$\Gamma=(D_\xi Z)\ln(B/B_0)/I_0$, depends on the birefringent disorder, the the initial signal
shape, as well as the details of the compensation and detection procedures, it is, however,
insensitive to the noise. Since OSRN is large, even weak disorder can generate strong increase in
the value of $B$. This is why a perturbative (with respect to $\bm h$) calculation of $\Gamma$
gives the most essential part of the PDF ${\cal S}(B)$ of $B$. If no compensation is applied, one
gets $\Gamma\sim H_3/b$, $b$ being the pulse width, and the initial signal is assumed to be
linearly polarized. In the simplest case of the ``setting the clock" compensation one arrives at
$\Gamma\sim(H_1^2+H_2^2)/b^2$. This yields the power-like tail of the PDF of $B$ \cite{03CCKLb}.
Higher-order compensation leads to $\Gamma\sim (H/b)^{p}$, where $p$ is an integer exceeding by
one the compensation degree provided no additional cancellations occur, and one gets the following
asymptotic expression (tail) for the PDF ${\cal S}(B)$ of $B$ \cite{03CCKLc}:
 \begin{equation}
 \ln{\cal S}= -\mu_p b^2
 [D_\xi Z\ln(B/B_0)/I_0]^{2/p}/[D_m Z],
 \label{korder} \end{equation}
where $\mu_p$ is a dimensional coefficient. Therefore, as anticipated, compensation suppresses
the PDF tail. However, applying high-order compensation is not too efficient, since the decrease
of ${\cal S}$ is mild as $p$ grows.

\begin{figure}
 \centerline{\includegraphics[width=0.4\textwidth]{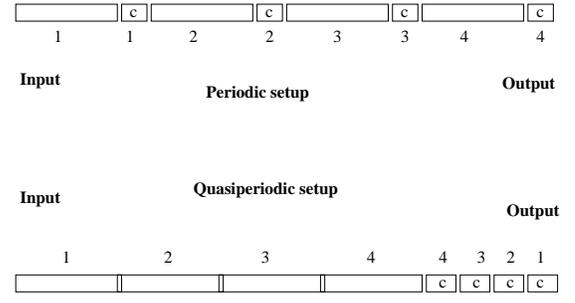}}
 \caption{Cartoon scheme of fiber-line elements
 installation correspondent to the periodic and
 quasi-periodic compensation strategies.}
 \label{periodic} \end{figure}

The main purpose of this letter is introducing more efficient compensation strategies with the
same number of compensating degrees of freedom. As a first example consider the following
``periodic" scheme. One divides the optical line into $N$ segments, each of the length $l=Z/N$,
and apply the first-order compensation at the end of each segment (as schematically shown in the
upper panel of Fig. \ref{periodic}, with ``c" denoting the compensating elements). The
noise-independent part of the compensated signal for the ``periodic compensation" strategy is
determined by
 \begin{eqnarray}
 {\cal K}_c\bm\varphi=\exp(i\eta\partial_t^2)
 {\cal K}_{1N}\hat U_N \dots
 {\cal K}_{11}\hat U_1
 \bm\varPsi_0(t) \,,
 \label{period} \\
 \hat U_n= T\!\exp\left[\int_{(n-1)l}^{nl}\!\!\!\mathrm dz\,
 h_j(z) \hat\sigma_j\partial_t\right],
 \label{tusp} \\
 {\cal K}_{1n}=\exp\left[-\int_{(n-1)l}^{nl}\!\!\! \mathrm dz\,
 h_j(z) \hat\sigma_j\partial_t\right] \,,
 \label{k1sp} \end{eqnarray}
where $\bm\varPsi_0(t)$ is the input signal profile, $\eta=\int_0^Z\!\!\mathrm dz\,d(z)$ is the
integral chromatic dispersion, and the ordered product on the rhs of Eq. (\ref{period}) is taken
over all the $N$ segments ($T\exp$ is the standard notation for the so-called ordered exponential).
The exponential factor ${\cal K}_{1n}$ represents the first-order compensation at the end of the
$n$-th segment.

This ``periodic" compensation is not too convenient since it requires installation of compensating
elements at multiple places along the fiber. However, one can naturally modify this scheme and
have the same compensating elements inserted subsequently but all at once at the fiber output as
it is shown in the lower panel of Fig. \ref{periodic}. If the disorder profile $h_j$ is known
(technically such end-point measurements are possible through the anti-Stokes refraction technique
\cite{99HGG}) one can have an end-point, but multiple, compensation as ${\cal K}_c=\prod{\cal
K}_{1n}$, leading to the following ``quasi-periodic" modification of Eq. (\ref{period}):
 \begin{equation}
 {\cal K}_c\bm\varphi=\exp(i\eta\partial_t^2)
 {\cal K}_{11}\dots{\cal K}_{1N}
 \hat U_N\dots \hat U_1 \bm\varPsi_0(t) \,.
 \label{quasi} \end{equation}
The natural idea behind this ``quasi-periodic" compensation is obvious: to construct (in the
compensating part) the best possible approximation (with the given number of the compensating
degrees of freedom) for the inverse of the ordered exponential $\hat U_N\dots \hat U_1$.

Note that the (quasi) periodic compensation does not influence the noise-dependent part of the
signal, i.e. ${\cal K}_c\bm\phi$ has the same correlation function (\ref{phiphi}) as $\bm\phi$.
Therefore, one arrives at the same expression $\ln(B/B_0)=\Gamma I_0/(D_\xi Z)$, with a new $\bm
h$-dependent factor $\Gamma$. Furthermore, in the region of our main interest $\Gamma$ can be
analyzed perturbatively, just as in \cite{03CCKLb,03CCKLc}. Expanding the factors in Eq.
(\ref{period}) up to the second order and making use of Eqs. (\ref{nnn},\ref{phiphi}) one derives:
 \begin{eqnarray}
 \Gamma\!\approx\!\frac{\mu'_2}{b^2}\sum_{n=1}^N\!
 \int_{a_n}^{nl}\!\!\!\! \mathrm dz\,
 \int_{a_n}^z\!\!\! \mathrm dz'\,
 [h_1(z) h_2(z')-h_2(z) h_1(z')],
 \label{second} \end{eqnarray}
where $a_n=(n-1)l$. Here, the dimensionless coefficient $\mu'_2$ is related to the output signal
chirp produced by an initial chirp and the nonzero integral chromatic dispersion $\eta$. As follows
from Eq. (\ref{quasi}), the same expression (\ref{second}) is obtained in the second order for the
quasi-periodic case. Substituting Eq. (\ref{second}) into the expression for $B$ and evaluating the
PDF of $B$, with the Gaussian statistics of $\bm h$ described by Eq. (\ref{hh}), leads to the
following expression for the tail of the PDF of $B$:
 \begin{equation}
 {\cal S}(B)\,\mathrm d B
 \sim\frac{B_0^\alpha\,\mathrm d B}
 {B^{1+\alpha}}, \qquad
 \alpha=\frac{N \pi D_\xi b^2}{2|\mu'_2| D_m I_0}.
 \label{powerN} \end{equation}
Eq. (\ref{powerN}) holds for $\ln(B/B_0)\gg \mu'_2 D_m I_0/[D_\xi b^2]$. The exponent $\alpha$ in
Eq. (\ref{powerN}) contains an additional factor $N$ compared to the expression for the first order
end-point compensation, i.e. the (quasi) periodic compensation makes the
tail of ${\cal S}(B)$ steeper. It is instructive to compare the outage probability for the
periodic case with the case of higher-order end-point compensation described by Eq.
(\ref{korder}). One finds that for higher-order compensation, i.e. when $N\sim \mu'_2
\ln(B_\ast/B_0)I_0/(D_\xi Z)$, the (quasi) periodic scheme becomes more efficient
compared to the
straight $N$-th order compensation scheme. If the output signal is not chirped,
$\mu'_2=0$ and the leading term in the expansion of $\Gamma$ in $\bm h/b$ is of the third order.
Additional filtering efforts can be made to enforce the output pulse symmetry under the $t\to-t$
transformation, thus removing the third-order term. Then the leading term in $\Gamma$ will be of
the fourth order in $\bm h/b$. Finally, even better compensation can be achieved if the standard
high-order compensation approach and the (quasi) periodic ones are combined, i.e. if in the
(quasi) periodic setting, considered above, one uses higher-order compensation instead of the
first order one. Formally, this hybrid case means that  the first-order compensation operators
${\cal K}_{1n}$ in Eqs. (\ref{period},\ref{quasi}) should be substituted by higher-order
compensation operators ${\cal K}_{cn}$. In the hybrid periodic case $\Gamma$ can be written as a
sum of $\Gamma_n\sim(\int_{a_n}^{ln}\mathrm dz\,\bm h/b)^p$, and, since $\bm h$ is
short-correlated, $\Gamma_n$ related to different segments are statistically independent. This
leads to the following expression for the PDF tail
 \begin{equation}
 \ln{\cal S}(B)\sim  -\mu_p N^{2(p-1)/p}
 \frac{[D_\xi Z\ln({B}/{B_0})b^p/I_0]^{2/p}}{D_m Z} \,.
 \label{powerk} \end{equation}
valid at $D_\xi Z/I_0\ln(B/B_0)\gg N^{1-p/2}(D_m Z/[\mu_p b^2])^{p/2}$.

Note, that an important computational step, leading to our major results in Eqs.
(\ref{powerN},\ref{powerk}), was evaluation of $\Gamma$ pertubatively in ${\bm h}$.
Besides, in the periodic case $\Gamma$ is a direct sum of each segment contribution $\Gamma_n$,
and the perturbative treatment applies separately to each $\Gamma_n$, requiring the
weakness of the PMD effect at each segment only, i.e. $D_m Z/N\ll b^2$. Therefore,
one concludes that even
an optical line with not really operable (without compensation) characteristics ($D_m Z$ which is
of the order or larger than $b^2$) can still be used for transmission if $N$ is sufficiently large.
Moreover, this observation on the applicability of Eqs. (\ref{powerN},\ref{powerk}) also extends
to the quasi-periodic case, in the sense that Eqs. (\ref{powerN},\ref{powerk}) provide an upper
bound for the PDF of BER. This is due to  an additional, oscillatory with $\bm h$,
suppression of $\Gamma_n$ in the quasi-periodic case vs periodic. This
suppression is especially important for segments strongly separated from their compensating
counter-segments.

For illustration purposes let us briefly discuss an example of a fiber line with typical bit error
probability, $B_0=10^{-12}$, and $\mu_2'=0.14$. Assume also that the PMD coefficient,
$k=\sqrt{12D_m}$, is $1.5$ $ps/\sqrt{km}$, the pulse width, $b=25$ $ps$, and fiber length
$Z=2,500$ $km$. Then the dimensionless parameter, $D_m Z/b^2$ measuring the relevant strength of
the PMD effect is $O(1)$, i.e. without compensation PMD effect is large, pulses are destroyed and
no successful transmission is possible: $S(B)=0(1)$ for any $B>B_0$. If, however, the (quasi)
periodic compensation with $N=10$ compensation units is utilitized the relevant strength of the
PMD effect is essentially reduced, so that $D_m Z/[b^2 N]\approx 0.1$, and $S(B)$ starts to decay
with $B$ at $B>B_0$. The system performance can be evaluated in terms of the outage probability
${\cal O}$,the probability for $B$ to be larger than $B_\ast$: ${\cal O}=\int^1_{B_\ast}\mathrm
dB\,{\cal S}(B)$. One derives from Eq. (\ref{powerN}) that in the (quasi) periodic case, ${\cal
O}\approx 0.06$ for $B_\ast=10^{-8}$ and ${\cal O}\approx 0.02$ for $B_\ast=10^{-6}$, i.e. the
system performance is essentially improved (to become not yet perfect but already satisfactory).

To conclude, in this letter we have proposed a (quasi) periodic compensation scheme which appears
to be a strong alternative to the standard higher-order compensations. The efficiency of the
scheme has been demonstrated. Even though technical implementation of this procedure needs an
expensive equipment, we anticipate that if this compensation technique is implemented the
reduction in the probability of extreme outages will guarantee an essential overall benefit.

The authors thank I. Gabitov for numerous valuable discussions.

\end{document}